# Towards Continuous Integration and Continuous Delivery in the Automotive Industry

**Abstract**

*Development cycles are getting shorter and Continuous Integration and Delivery are being established in the automotive industry. We give an overview of the peculiarities in an automotive deployment pipeline, introduce technologies used and analyze Tesla's deliveries as a state-of-the-art showcase.*

**Introduction**

The revolution in digital systems has a massive effect on our daily life and on all branches of industry. Industries traditionally dominated by mechanical engineering are now shifting to be software-driven, as can be observed in the transition of the automotive industry.

Customers today are used to getting the latest updates automatically, instantly and free. This development begun with agile methods gaining popularity in application software development, thus allowing fast release-cycles. "Fail Fast" has become the driving motive of innovation-generating Silicon Valley companies. Today, *Continuous Delivery* is state-of-the-art in certain application domains and enables software developers to provide a new release to a broad base of customers at the push of a button. This allows entirely new concepts of development and acceptance testing. Today, a new function can be provided to a limited group of users to receive instant customer feedback.

But what if the customers are drivers instead of mobile phone users? The major difference here is that a car is a safety-critical system and faults in software may lead to injury or death. So if a car manufacturer intends to "Fail Fast", they will have to do so before they deliver and thus perform thorough and automatic checks of their software if they intend to deliver continuously.

This is becoming ever more important with the industry starting to deploy autonomous driving. Customers essentially put their life fully in the hands of some piece of software if they trust an autonomous car to get them to their destination safely. At the same time, they expect that piece of software to constantly be at the very latest edge of technology. These antithetic requirements impose a tremendous challenge to the automotive industry more than any other branch, because it is the only cyber-physical system produced in large series today.

This article shall give an insight into how the automotive industry is trying to overcome these challenges today, what technologies are used and what limitations still exist today.

**Deployments in Automotive Software – Distributed, Embedded, Enterprise, Safety-Critical Systems**

Before any software can be delivered to the customer, a number of steps have to be performed which includes compiling and assembling the final product, but also testing on different levels.
These steps can be summarized as subsequent stages of a deployment pipeline [1]. The first stage summarizes all the steps that can be performed automatically in a Continuous Integration matter, while the second stage is formed of acceptance tests that have longer running times. The final stage is a typical release along with a User Acceptance Test. Each stage is triggered only when the preceding stage was passed successfully.

Thus, when trying to deliver continuously, the goal has to be to perform stages two and three as quickly as the first stage (or at least nearly so). Otherwise, the cycle in which software changes and new releases are produced outpace the release process such that builds need to be grouped to a release and the whole idea of continuous deliveries disappears.

Indeed, technologies and tools from the DevOps Movements, such as Docker (https://www.docker.com/) or Puppet (https://puppet.com/) aim at automatizing the release and enable companies such as Amazon and Facebook to continuously deploy their latest builds on their production environments.

But what about companies like Tesla, which produce highly embedded and distributed systems? The automotive industry has entirely different structure, processes and requirements. So what does a (continuous) delivery pipeline look like in this domain? The first thing that needs to be considered in order to understand automotive release processes is the hardware- and software-architecture of a vehicle. Modern cars have up to 100 individual ECUs with different purposes that interact to implement a complex function like an ADAS (Advanced Driver's Assistance System). A schematic visualization can be seen in Figure 1. In order to reduce the load on the communication channels, certain pre-processing is already performed by the sensoric units themselves: For example, an ADAS ECU receives the data about the surroundings in the form of a list of objects that was generated from raw data by the camera and

---

**The origin of Continuous Software Engineering**

With agile software development becoming state-of-the-art, long integration cycles were obsolete and even obstructive. In 1991, the term "Continuous Integration" was first used by Grady Booch [1] to describe an effective, iterative way of building software. The technique was quickly adopted into the set of techniques used in Extreme Programming and detailed guidelines were summed up by Fowler in 2006 [2]: With a high degree of automation, fast integration tests and a single source-repository on which every single commit is stored, it is possible to set up a tool-supported pipeline that allows to create a stable build with the push of a button. With such capabilities, every commit is immediately followed by a full build in order to quickly detect errors and constantly have a stable build.

Fowler already stated that continuous builds should be deployed to production environments. This is the logical next step: If you continuously generate the latest, stable build, you only need to take a few more steps to actually get the software to the customer. Notably, these steps are acceptance tests of the full product, which can be automated to some extent. Adopting this is called Continuous Delivery or Continuous Deployment with the former term experiencing greater popularity in today's large DevOps-Movement.

While continuous integration is standard practice in many software projects and continuous delivery has been adopted by many large companies, most notably Facebook and Amazon, more and more aspects of software development are considered in a "continuous" process, to the extent of including even remotely related processes such as human resources [3]. These notions are summed up in the field of "Continuous Software Engineering" [4].

[1] Booch, Grady. *"Object oriented design with applications."* Redwood City." (1991)

[2] Fowler, Martin, and Matthew Foemmel. *"Continuous integration." Thought-Works)* http://www. thoughtworks. com/Continuous Integration. pdf *(2006).*

[3] Fitzgerald, Brian, and Klaas-Jan Stol. *"Continuous software engineering and beyond: trends and challenges."* Proceedings of the 1st International Workshop on Rapid Continuous Software Engineering. ACM, 2014

[4] Bosch, Jan, ed. *Continuous Software Engineering*. Springer, 2014.

radar sensors in advance. Thus, automotive systems today are not only embedded, but distributed embedded systems.

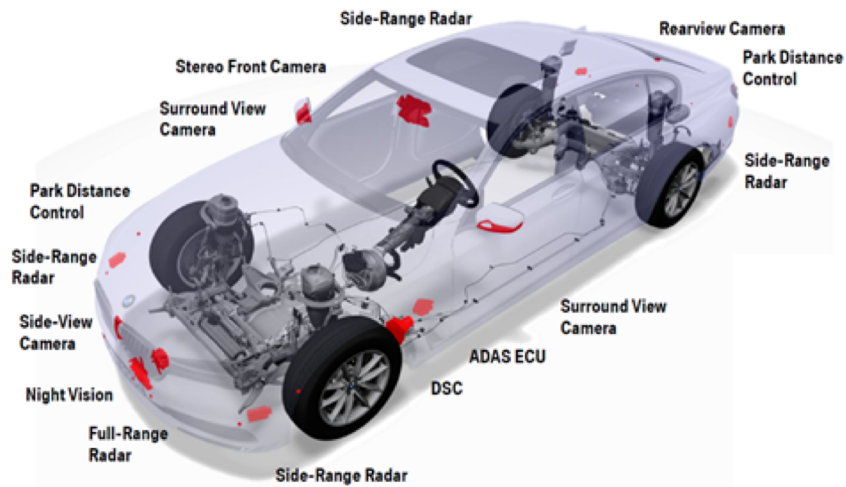

Figure 1: Schematic View of Sensors, Actuators and Processors for ADAS in a modern Vehicle

Due to the organizational structure of OEMs, software teams are equally distributed and heterogeneous. ECUs are usually developed and produced by a contractor, including all software. However, hybrid structures are also possible: parts or all of the application software may be supplied by the OEM or other contractors and the deployment on the ECU may be done by another contractor, a subcontractor or even the OEM again. This implies that a delivery may be triggered by many different sources that have already passed through their own deployment pipeline.

Finally, a car is a safety-critical system, which means that functional safety checks have to be performed before any deployment to the productive environment. Typically, analysis methods like FMEA [2] or STPA [3] are used. They involve a safety analyst identifying possible hazardous scenarios and testing them against the release candidate.

**Stages of the Automotive Continuous Delivery Pipeline**

These requirements lead to somewhat different stages in the automotive industry that need to be performed redundantly in parallel by different organizations. As shown in Figure 2, the starting point is always the commit of some source code in a single ECU. What follows immediately is a standard Continuous Integration pipeline including static code analysis, compilation, unit and integration tests. Since one ECU may contain one or more libraries of application software and always has a separate operating system that supplies an abstraction layer for basic functions such as scheduling or communications, further integration tests have to be performed. Like in large enterprise software, a build needs to be triggered when one of the dependencies changed [4].

The different libraries first need to be configured and linked to the operating system. This is typically a manual task and supported by specialized tools such as Vector's DaVinci Configurator (https://vector.com/vi_davinci_configurator_pro_en.html ), which – amongst others – offers visualization of involved libraries and their interfaces. While this step today often requires expert knowledge, this can be derived from architectural information and can be automatized when a continuous delivery pipeline is set up. The result of this step is a fully functional container that can be flashed on any compatible hardware. This is indeed very comparable to the containerization known from DevOps.

The next step involves Integration Tests performed on a single ECU. These tests are supposed to ensure the correct functionality of the isolated control unit and is thus limited to interface tests. These tests are eventually carried out on custom Hardware-in-the-Loop or Open-Loop test benches. Many different build management systems, from the well-known Jenkins to proprietary custom developments with sophisticated test selection methods, are in use to trigger the tests. The concrete test execution is performed by standard tools like ECU-Test (https://www.tracetronic.com/products/ecu-test/ ) or CanOE (http://vector.com/vi_canoe_en.html ). These tools are connected to the bus systems and can monitor, interpret and manipulate the signals when given the according architectural information. They offer interfaces to specify a desired behaviour, e.g. the simulated input of a signal from a communicating ECU along with pass/fail criteria, such as the expectation of a certain signal being sent within a given time.

If all ECU integration tests passed, the compiled software containers are committed to a central repository. This can be compared to a "commit" into a source repository, with the difference that binary artifacts are checked in and this is the second pipeline that is being activated. However, this commit triggers the deployment of software first on integration test benches and later on test vehicles.

The test benches used vary in complexity and testing goal. When it comes to functional integration, however, most test benches are designed so that they contain all ECUs that implement a certain number of functions along with a complex Hardware-in-the-Loop-Simulation for the environment and further ECUs. The tests for these functions will be executed on this type of test bench.

However, the effort in this stage is gigantic. A modern vehicle contains a large number of functions. ISO26262 demands test coverage for each requirement of a function, which results in a number of several 10.000s of integration tests for the entire vehicle. And since they are executed on the target hardware, they have to run in realtime which leads to an average execution time of several minutes per test.

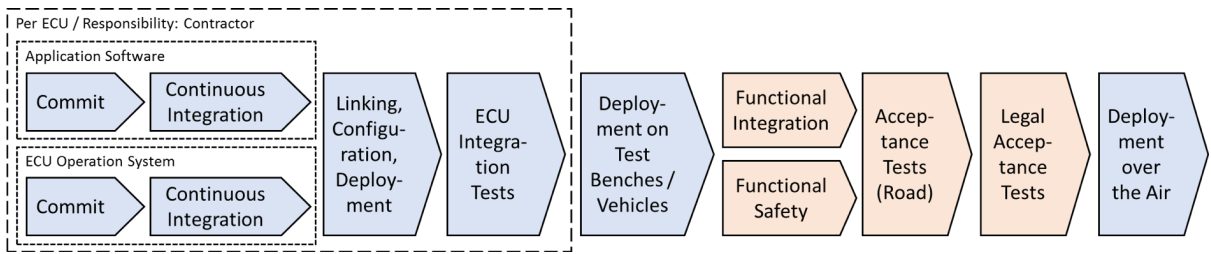

Figure 2: The Delivery Pipeline in Automotive Software

> **Hardware-in-the-Loop and Open-Loop**
>
> Embedded Systems or Cyber-Physical Systems are unique in their property that they directly interact with the physical world. As such they require input and provide output which cannot always be generated or checked easily when testing software.
>
> To avoid the necessity to manipulate physics, complex systems like ADAS are often tested in simulated environments [1]. In this approach, sensors are replaced with a computer that provides the exact signals the sensor would send under the desired physical condition. For actors respectively a computer is connected that can translate the outgoing signals to a hypothetical physical action.
>
> Open-Loop and Hardware-in-the-Loop are two oppositional approaches. The first consider system interactions individually and simulate only one action and the corresponding reaction. This makes it lightweight and easy to set up, but has limitations when complex staged interactions like in ADAS need to be tested. Hardware-in-the-Loop-systems on the other hand replace the entire environment with extremely complex mathematical models and can simulate an entire control loop: It is able to calculate the effects of the system's action on the environment and can thus properly react even on consecutive system actions.
>
> Apart from hardware, also software and even system models can be put into the loop and are used for more lightweight tests in earlier phases.
>
> [1] O. Gietelink, J. Ploeg, B. De Schutter, and M. Verhaegen, *"Development of advanced driver assistance systems with vehicle hardware-in-the-loop simulations"*, Vehicle System Dynamics, vol. 44, no. 7, pp. 569–590, July 2006.

Obviously, this number needs to be reduced and a test suite for the exact change has to be tailored. In a typical manual functional integration process, expert knowledge is employed to analyze which sub-function is tested in which test case. In a Continuous Scenario however, the expert needs to be replaced with heuristics. In Software Engineering, many methods for test selection have been proposed, but since the source code is not available, most of these do not apply. In this case, test selection methods that analyze a test suite regarding the communication paths involved show great potential: Tests are only executed if the signals manipulated and checked are actually being processed by the changed piece of software [6].

These tests are executed on so-called "Test Farms". These are conglomerates of similar test benches with different configurations to allow parallel execution. In some cases, tests have requirements on certain configurations (e.g. a test case is designed for a certain powertrain system). Such a test farm has to be controlled by a central server system. Apart from knowing the status of the test benches and the tests queued, such a system must contain a component for load-balancing to ensure maximum parallelization and that every test is executed on a test bench with the right configuration. Some test bench designs allow reconfiguration during runtime, for example by having redundant ECU variants that can be connected using a relay. In this case, the task of the load-balancer becomes a complex scheduling problem. This is a task currently only performed by custom server systems [7].

Testing for functional safety is a base requirement to allow the software to be flashed on a vehicle that enters the road. As mentioned before, the system's requirements and architecture are analyzed using a method like FMEA or STPA. This results in test cases that are executed on the target hardware. This can be integrated into a Continuous pipeline just like functional integration tests, because the same technologies are used. The difference is though, that the safety analysis has to be performed in a continuous matter, too, and there is no attempt in the literature yet to do this.

The pipeline stages as described so far can be integrated into a continuous pipeline with more or less effort and do not hinder deliveries within a day or night if sufficient resources and proper and efficient tooling is available. This is different with the two following steps, however: Acceptance tests on the road will always stay a manual action. Many of these tests can be automated and executed on test farms, potentially reducing this process to several days. Still, this causes a delay and considering several commits per day, not every build will be delivered in a "continuous" pipeline that includes this stage.

The final stage, deployment using over-the-air technology, is well-tested today as Tesla has shown. With high-speed mobile data standards such as 4G, updating large amounts of software even in remote areas is not a problem anymore. The only problem at this stage is that a car might not be receiving service for an extended period of time. In that case, deliveries cannot be made continuously, but this problem arises only in very remote areas.

**An Outlook to Palo Alto**

As a Silicon Valley company, Tesla aims to put their innovations on the market as quickly as possible. In fact, they are the only automobile manufacturer credited with Continuous Delivery [8]. But the mechanisms behind their deliveries are intransparent. According to the Tesla forums, users appear to be confused about whether or not their car has received the latest update yet [9]. The same update version appears to be deployed on different cars at different dates.

The publicly available database "Tesla Firmware Upgrade Tracker" (http://ev-fw.com/ ), which enables Tesla users to upload data about updates their car received, tracked over 1000 cars with over 5000 singular updates over the course of firmware version 7.1.

Figure 3 shows four representative builds and the number of cars on which they were deployed each day. The first and most obvious observation is that a build has some form of "lifecycle" in which it is gradually deployed until it becomes outdated and that these lifecycles overlap, sometimes to a large extent. It should be noted that overlapping builds are often deployed on the same car one after another, so it is safe to assume that not every build can be "skipped" when deploying.

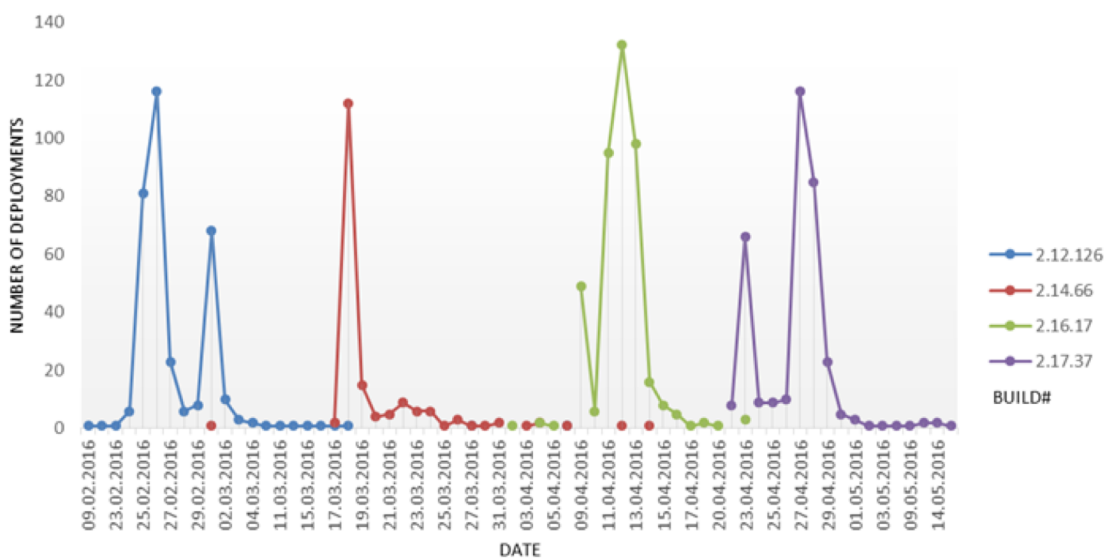

Figure 3: Selected Tesla Builds and the number of cars they were deployed on each day.

The "lifecycle" can be described in three phases:
- The "release date" is the first date on which a build is available to a broad public and many cars will be updated on this day.
- The "ramp-up-phase" typically consists of a few days following the release date. The vast majority of cars that require the update will be updated during this time. The delay is most likely caused by the availability of a wireless data service.
- During the "fade-away-phase", which can sometimes take several weeks to months, every other day a few cars receive the update. The reason for the length of this phase is unclear, but could be caused by a combination of vehicles not receiving wireless data for an extended time, limited time-slots for updates per day and dependencies on other updates that have to take place first.

This lifecycle can be observed in Figure 4, which displays the most frequently deployed updates in August and September 2016 as a heatmap. In this diagram, however, another anomaly can be observed. Almost every major build is deployed a few days prior to the release date, in some cases up to four weeks earlier. It is highly unlikely that this anomaly is caused by errors in the vehicles, such as update times being reported inaccurately, because it has been observed so regularly. 15 out of the 26 builds in Tesla's Firmware version 7.1 that have been reported on more than 50 cars were reported more than one day before the release date (note that tesla does not provide official release dates).

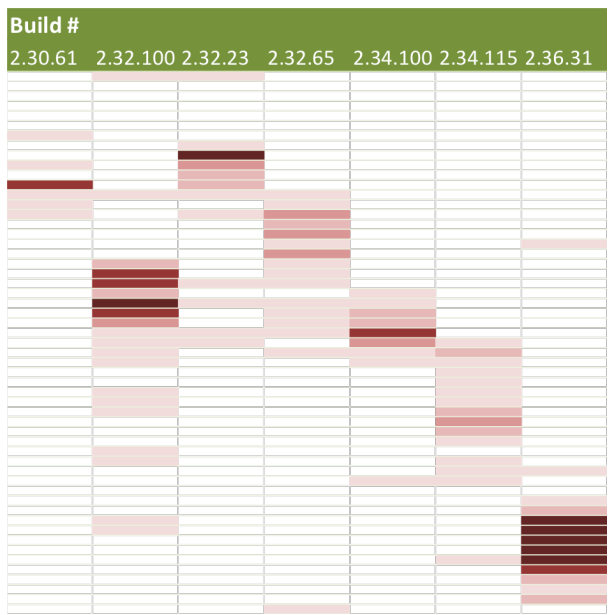

Figure 4: Most frequently deployed builds of Tesla Firmware 7.1 in 08/16 to 09/16.

Each row designates one day, the color indicates the number of cars that received the update on that particular day.

Furthermore, there is no connection between individual cars or models and the occurrence of such "early deployments". A single car may receive one update early, the next couple of updates on the release date or in the ramp-up-phase and some others even later. We could not observe that a single car has received an early deployment twice. What we observed here can be described as a form of the canary release pattern. [1]

While we can only speculate what the reason for this might be, this is clear evidence that Tesla deploys their software indeed continuously, but with a delay of up to four weeks or more, assuming that a build is a designated and most importantly fixed revision of Tesla's source code.

**Continuous Delivery: Way to go!**

A Continuous Delivery pipeline as complex and costly as this could easily explain a delay of several weeks from the commit of software to the final deployment in the production environment. The pipeline contains the full Continuous Delivery pipeline as known from other domains, but has to be passed redundantly and for each ECU individually. While on this level familiar technologies can be used and swift processing is easily achieved, the pipeline is only the first step towards a deployment in a complex, embedded, distributed, safety-critical system. All these properties of a vehicle impose additional requirements on the deployment pipeline that take their time.

Most significantly, necessary manual steps such as Acceptance Tests and Legal Approval can possibly delay the deployment of a new software. Errors found in these steps require a lengthy analysis before a fixed version can be sent into the pipeline.

Yet we observe an enormous number and frequency of releases on Tesla's vehicles. This indicates that from a technical point of view, most of the problems with Continuous Delivery in the automotive industry can be and are being solved. If the pace of past years' developments can be sustained, a swift and full Continuous Integration Pipeline will be established throughout the industry within the next years.

**Authors**

Sebastian Vöst is a PhD Student at the department of Software Integration at BMW Group. Contact him at Sebastian.Voest@bmw.de.

Stefan Wagner is a Professor at University of Stuttgart and head of the department for Software Engineering. Contact him at Stefan.Wagner@informatik.uni-stuttgart.de